\documentclass[prb,aps,12pt]{revtex4}
\usepackage{amsmath}
\usepackage{amssymb}
\usepackage{graphicx}

\begin{document}

\title{Phase diagram and macroscopic ground state degeneracy of frustrated spin-1/2 anisotropic Heisenberg model on diamond-decorated lattices}
\author{D.~V.~Dmitriev}
\email{dmitriev@deom.chph.ras.ru}
\author{V.~Ya.~Krivnov}
\author{O.~A.~Vasilyev}
\affiliation{Institute of Biochemical Physics of RAS, Kosygin str. 4, 119334, Moscow,
Russia.}
\date{}

\begin{abstract}
We study the ground state properties of the anisotropic spin-1/2
Heisenberg model on lattices built from ideal diamond units with
competing ferro- and antiferromagnetic interactions. The study
covers the one-dimensional diamond chain and its two- and
three-dimensional generalizations. The ground-state phase diagram
contains four distinct phases: ferromagnetic (F), critical (C),
monomer-dimer (MD), and tetramer-dimer (TD), which converge at a
quadruple point. We demonstrate the presence of macroscopic
ground-state degeneracy and corresponding residual entropy, which
is maximal at the quadruple point and also extends throughout the
MD phase and its boundaries with TD and F phases. For the diamond
chain, we derive exact degeneracies, while for higher-dimensional
lattices, we map the problem onto a bond percolation model or used
transfer-matrix approach, enabling the numerical computation of
the ground state degeneracy.
\end{abstract}

\maketitle

\section{Introduction}

Quantum magnets on geometrically frustrated lattices have been
extensively studied in recent years \cite{diep, diep2, mila}. A
noticeable class of these systems involves lattices with magnetic
ions located at the vertices of connected triangles. For specific
relations between exchange interactions, these systems exhibit a
macroscopically degenerate ground state. As examples of such
systems are the spin models with dispersionless (flat) one-magnon
band. Frustrated spin systems with flat-band physics have attracted
significant attention since the early 2000s. This interest was
stimulated by the seminal works of J.Richter and co-workers \cite{schulenburg2002macroscopic, flat}. Flat band in the
one-magnon spectrum means that magnons
localized within a small part of the lattice, trapping cell. This
phenomenon has been observed in a broad class of highly frustrated
antiferromagnetic spin systems \cite{derzhko2007universal,
zhitomirsky2005high}. The
localization of one-magnon states forms the basis for constructing
multi-magnon states, because states consisting of isolated
(non-overlapping) localized magnons are exact eigenstates.

In antiferromagnetic flat-band models, localized states constitute
the ground state manifold in the saturation magnetic field,
leading to an exponentially growing degeneracy in the
thermodynamic limit and residual entropy. The ground state
properties and low-temperature thermodynamics of these models have
been extensively studied, revealing intriguing features such as a
zero-temperature magnetization plateau, an extra low-temperature
peak in the specific heat, and an enhanced magnetocaloric effect
\cite{flat, shulen, zhitomirsky2005high, schmidt, honecker, hon2,
Derzhko, derzhko2006universal, richter2018thermodynamic,
zhitomirsky2003enhanced}.

Another class of frustrated quantum models with a one-magnon
flat band involves systems with competing ferro- and
antiferromagnetic interactions (F-AF models). The zero-temperature
phase diagram of these models exhibits different phases depending
on the ratio of ferromagnetic to antiferromagnetic interactions.
At the critical value of this ratio, corresponding to a phase
boundary (quantum critical point), the model exhibits a
macroscopically degenerate ground state. Examples include the
delta-chain at the critical value of the frustration parameter
\cite{zhitomir, Derzhko, KDNDR, DKRS, DKRS2, DKRS3} and its
two-dimensional generalizations on Tasaki and Kagome lattices
\cite{anis3}. In contrast to antiferromagnetic models, F-AF
systems support additional magnon complexes, which are exact
ground states at the critical frustration parameter. This leads to
macroscopic ground state degeneracy at zero magnetic field and a
higher residual entropy compared to antiferromagnetic models. The
residual entropy in zero magnetic field enhances magnetic cooling,
which is of practical importance.

Recently, further examples of frustrated F-AF spin models with
flat bands and macroscopic ground state degeneracy were studied
in \cite{diamond1d,DKV2D}. One of them is the spin-$\frac{1}{2}$
Heisenberg chain of distorted diamond units. (A diamond unit with
two different exchange interactions is referred to as an `ideal
diamond', while a unit with three different interactions is
called a `distorted diamond'). It was shown in \cite{diamond1d}
that this model has flat bands and a macroscopically degenerate
ground state for specific relations between exchange interactions.
Remarkably, this model features not only a one-magnon flat band
but also two- and three-magnon dispersionless bands, with the
corresponding localized multi-magnon states. Another model of this
kind studied in \cite{DKV2D}, is the spin-$\frac{1}{2}$ Heisenberg
model on a diamond-decorated square (cubic) lattice, where the
bonds in the square (cubic) lattice are replaced by diamonds. This
model with distorted diamond units supports up to five (seven)
localized magnon states within the trapping cell, all of which are
confirmed as exact ground states \cite{DKV2D}. All these states
belong to the ground state manifold, leading to an exponential
increase in ground state degeneracy compared to models with only
one-magnon localized states.

Systems with frustration-affected diamond units have attracted
significant attention both experimentally and theoretically.
Several recently synthesized compounds exhibit competing
ferromagnetic (F) and antiferromagnetic (AF) interactions within
the diamond unit, such as $K_{3}Cu_{3}AlO_{2}(SO_{4})_{4}$
(alumuklyuchavskite) \cite{alum} and $K_{2}Cu_{3}(MoO_{4})_{4}$
\cite{PhysRevB.111.144420}, as well as $Fe-Cu$ bimetallic magnetic
compounds with a diamond-decorated honeycomb structure
\cite{HONG20043271}.

Models composed of ideal diamond units with antiferromagnetic
interactions are of great interest and have been intensively
studied  \cite{takano1996ground, morita2016exact, hirose2016exact,
hirose2018ground, caci2023phases, karl2024thermodynamic}. These
one-dimensional and two-dimensional systems with isotropic
antiferromagnetic exchange interactions exhibit three types of
ground state phases, including the Lieb-Mattis ferrimagnet, a
monomer-dimer phase, and a tetramer-dimer phase, the latter two
possessing macroscopic ground state degeneracy.

While the ground state degeneracy of the spin models with
distorted diamonds arises from localized many-magnon states, such
states are absent in models with ideal diamonds. However, an
alternative route to the macroscopic degeneracy of the ground
state exists: if model eigenstates are direct products of
eigenstates of isolated clusters and lowest energies per spin of
different isolated clusters are equal, the total number of ground
state grows exponentially with the system size. As shown in
\cite{diamond1d,DKV2D} this scenario is realized in F-AF models
with the ideal diamond unit for special relations between F and AF
interactions, where the ground state consists of ferromagnetic
clusters embedded in the background of diamonds with diagonal
singlets.

In this paper, we extend our previous investigation of the
isotropic model \cite{diamond1d,DKV2D} to the anisotropic F-AF
models with ideal diamonds. Our particular attention is focused on
the parameter region favoring the macroscopic ground state
degeneracy. The anisotropic models, governed by three parameters,
allow us to explore a ground state phase diagram and identify the
parameter relations leading to macroscopic degeneracy. For the
F-AF diamond chain, we obtain the ground state degeneracy in
analytical form. For diamond decorated lattices, counting the
number of ground states maps to a bond percolation problem, which
we solve numerically using Monte Carlo techniques.

The paper is organized as follows. In Section II, we introduce the
Hamiltonian of the anisotropic spin-$\frac{1}{2}$ F-AF Heisenberg
chain with ideal diamonds and analyze its ground state phase
diagram. We discuss the properties of different phases and
determine their degeneracies, identifying the conditions for the
maximal degeneracy. In Section III, we consider the anisotropic
spin-$\frac{1}{2}$ F-AF Heisenberg model with ideal diamonds on
two- and three-dimensional diamond decorated lattices and
calculate the ground state degeneracy via a mapping to a
percolation problem. Finally, in the concluding Section, we
summarize our key findings.

\section{Ideal anisotropic diamond chain}

\begin{figure}[tbp]
\includegraphics[width=3in,angle=0]{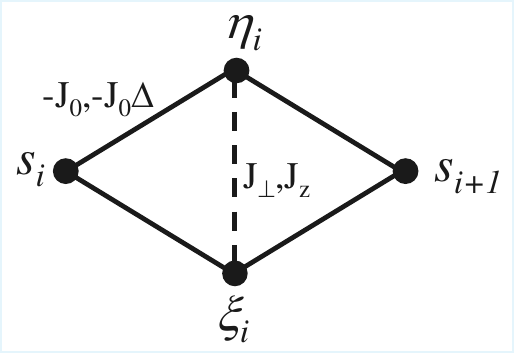}
\caption{Ideal diamond unit described by Hamiltonian
(\ref{h}). The solid lines denote the anisotropic ferromagnetic bonds, the dashed line represents the anisotropic antiferromagnetic diagonal bond.} \label{Fig_diamond}
\end{figure}

In this Section we study the F-AF anisotropic diamond chain. The
Hamiltonian of this chain can be represented as a sum of local
Hamiltonians
\begin{equation}
\hat{H}=\sum_{i=1}^{n}\hat{H}_{i}  \label{H}
\end{equation}%
where $\hat{H}_{i}$ is the local Hamiltonian of $i$-th diamond,
shown in Fig.\ref{Fig_diamond}, which can be written in the form:
\begin{eqnarray}
\hat{H}_{i}
&=&-J_0[(s_{i}^{x}+s_{i+1}^{x})L_{i}^{x}+(s_{i}^{y}+s_{i+1}^{y})L_{i}^{y}+%
\Delta (s_{i}^{z}+s_{i+1}^{z})L_{i}^{z}]  \notag \\
&&+\frac{J_{\perp }}{2}\mathbf{L}_{i}^{2}+\frac{J_{z}-J_{\perp }}{2}%
(L_{i}^{z})^{2}+J_0\Delta -\frac{1}{2}(J_{\perp }+J_{z})  \label{h}
\end{eqnarray}

Here $\mathbf{L}_{i}=\mathbf{\xi }_{i}+\mathbf{\eta }_{i}$ is
the composite spin on the diamond diagonal with quantum number
$L_{i}=0$ or $L_{i}=1$, $n$ is the number of diamonds. Therefore, the anisotropic interaction between diagonal spins $\mathbf{\xi }_{i}$ and $\mathbf{\eta }_{i}$ is written in Eq.(\ref{h})
in terms of composite spin $\mathbf{L}_{i}$ as
\begin{equation}
J_{\perp }(\xi_{i}^{x} \eta_{i}^{x} + \xi_{i}^{y} \eta_{i}^{y})
+ J_z (\xi_{i}^{z} \eta_{i}^{z} - \frac{1}{4})
= \frac{J_{\perp }}{2}\mathbf{L}_{i}^{2}+\frac{J_{z}-J_{\perp }}{2}(L_{i}^{z})^{2}
-\frac{J_{\perp }+J_{z}}{2}
\end{equation}%
The
interaction $J_0$ between the central spins $s_{i}, s_{i+1}$ and
diagonal spin $\mathbf{L}_{i}$ is ferromagnetic, and we take $J_0$ as
an energy unit. The constants in (\ref{h}) are chosen so that the
energy of the fully polarized states is zero, and the periodic
boundary conditions are imposed.

The eigenstates of the local Hamiltonian $\hat{H}_{i}$ are
classified by the quantum number $L_{i}$. The state with $L_{i}=0$
is the singlet located on the diagonal
\begin{equation}
\hat{\varphi}_{i}\left\vert F\right\rangle =(\eta _{i}^{-}-\xi
_{i}^{-})\left\vert F\right\rangle   \label{dstate}
\end{equation}%
where $\eta _{i}^{-}$ and $\xi _{i}^{-}$ are spin-lowering
operators and $\left\vert F\right\rangle $ is the ferromagnetic
state with all spins up. We denote (\ref{dstate}) as the `dimer'
state. The energy of this dimer state is
\begin{equation}
\varepsilon _{dimer}=\Delta -\frac{1}{2}(J_{\perp }+J_{z})  \label{dimer}
\end{equation}

The dimer state is an exact eigenstate of both local and total
Hamiltonians because $L_{i}=0$ decouples spins $s_{i}$ and
$s_{i+1}$. Therefore, the states
$s_{i}^{-}\hat{\varphi}_{i}\left\vert F\right\rangle ,$
$s_{i+1}^{-}\hat{\varphi}_{i}\left\vert F\right\rangle $ and
$s_{i}^{-}s_{i+1}^{-}\hat{\varphi}_{i}\left\vert F\right\rangle $
are exact ones with the same energy $\varepsilon _{dimer}$.

The energy of the states of $\hat{H}_{i}$ with $L_{i}=1$ and the total
spin projection $S^{z}=\pm 2$ is $\varepsilon ^{(2)}=0$. It turns
out that for $L_{i}=1$, the state with minimal energy is given by
either the polarized state with $\varepsilon ^{(2)}=0$, or by the
lowest state with $S^{z}=0$ (`tetramer' state) and the energy
\begin{equation}
\varepsilon _t =\frac{3}{2}\Delta +\frac{J_{\perp }-J_{z}}{4}-
\frac{1}{2}\sqrt{\left( \Delta -\frac{J_{\perp }-J_{z}}{4}\right)
^{2}+8} \label{ESz0}
\end{equation}

As will be demonstrated in the following, the ground state properties of
model (\ref{H}) depend on the relative values of $\varepsilon
_{dimer}$, $\varepsilon _{t}$, and the energy of the ferromagnetic
state, which is normalized to zero.

\subsection{Ground state phase diagram for diamond model with $J_{z}=J_{\perp }=J$}

We begin by examining the ground state phase diagram of model
(\ref{H}) in the case of isotropic diagonal interactions $J_{\perp
}=J_{z}=J$.

All eigenstates of (\ref{H}) are described by a definite
configuration of singlets, $L_{i}=0$, located in diamonds
$\{i_{1},i_{2},\ldots i_{k}\}$. Effectively, each singlet
$L_{i}=0$ cuts the chain and creates an open boundary in this
place. Therefore, for the given configuration of singlets
$\{i_{1},i_{2},\ldots i_{k}\}$, there are $k$ sections of open
mixed spin chains of different lengths $m_{j}=(i_{j+1}-i_{j}-1)$,
located between singlets $i_{j}$ and $i_{j+1}$ and comprised of
alternating $m_{j}+1$ spins-$\frac{1}{2}$ and $m_{j}$ spins-$1$
(the anisotropic mixed spin-$(\frac{1}{2},1)$ chain). The lowest
energy corresponding to the configuration $\{i_{1},i_{2},\ldots
i_{k}\}$ can be written as
\begin{equation}
E\{i_{1},i_{2},\ldots i_{k}\}=k\varepsilon
_{dimer}+\sum_{j=1}^{k}\varepsilon _{0}(m_{j})  \label{Em}
\end{equation}%
where the energy of the dimer state (\ref{dimer}) in the case
$J_{z}=J_{\perp }=J$ is
\begin{equation}
\varepsilon _{dimer}=\Delta -J  \label{Edimer}
\end{equation}%
and $\varepsilon _{0}(m_{j})$ is the lowest energy of the open mixed spin
chain of length $m_{j}$.

The ground state configuration, denoted by the collection of
singlet pairs $\{i_{1},i_{2},\ldots i_{k}\}$, depends on the model
parameters $\Delta $ and $J$. All eigenstates of (\ref{h}) consist
of the eigenstates of the open mixed spin $(\frac{1}{2},1)$ segments of
various lengths and of the singlet states. For $J<\Delta$, the
dimer energy is positive, $\varepsilon _{dimer}>0$, and the ground
state of (\ref{H}) is that of the spin chain of alternating spins
$s=\frac{1}{2}$ and spins $L=1$ (mixed spin-$(\frac{1}{2},1)$
chain). As was shown in \cite{alcaraz1997critical} the ground
state of this anisotropic mixed spin chain displays distinct
features depending on the value of $\Delta $ (similar to the
conventional XXZ chain).

For $\Delta >1$, the ground state takes the form of a two-fold
degenerate, fully-polarized configuration with maximal total spin
projections $S_{tot}^{z}=\pm S_{\max }^{z}$. Exactly at the
isotropic point ($\Delta =1$), the ground state consists of a
manifold of ferromagnetic states, including all allowed
multiplets, ranging from $-S_{\max }^{z}$ to $S_{\max }^{z}$. When
$\Delta <1$, the ground state lies within the non-magnetic spin
sector $S_{tot}^{z}=0$, indicative of a critical (C) phase, similar to
that of the easy-plane spin-$\frac{1}{2}$ Heisenberg XXZ chain
\cite{alcaraz1997critical}.

Switching to case $J>\Delta $ leads to a change in the sign of
the dimer energy ($\varepsilon _{dimer}<0$), introducing new
ground state properties dependent on the value of $\Delta $. If
$\Delta \geq 1$, the lowest energy of the open mixed spin
$(\frac{1}{2},1)$ segments is positive, so the ground state
is a product of singlets on all diagonals (all $L_{i}=0$). All
central spins $\mathbf{s}_{i}$ effectively decouple, which
produces $2^{n}$ degeneracy of the ground state. This is the so
called monomer-dimer (MD) phase. Alternatively, for $J>\Delta $
and $\Delta <1$, both the dimer energy and the lowest eigenvalues
of open mixed finite chains are negative, and the ground state of
model (\ref{h}) depends on the relation between them.

Based on extensive numerical computations and variational
estimations, we establish that the inequality $\varepsilon
_{0}(1)<\varepsilon _{0}(m)/m$ holds true for $\Delta <1$, so that
the energy (\ref{Em}) is minimal when all $m_{i}=1$. This means that
two tetramers effectively repel each other. (Notably, this
phenomenon persists as a universal feature for two- and
three-dimensional systems with spin $s=1/2$). Consequently, the
emergence of a tetramer-dimer (TD) phase - a configuration marked
by a two-fold degenerate ground state involving alternating
singlets and triplets along the diagonals, $\{1,3,5\ldots \}$ and
$\{2,4,6\ldots \}$, - is feasible in certain region of the model
parameters.

\begin{figure}[tbp]
\includegraphics[width=5in,angle=0]{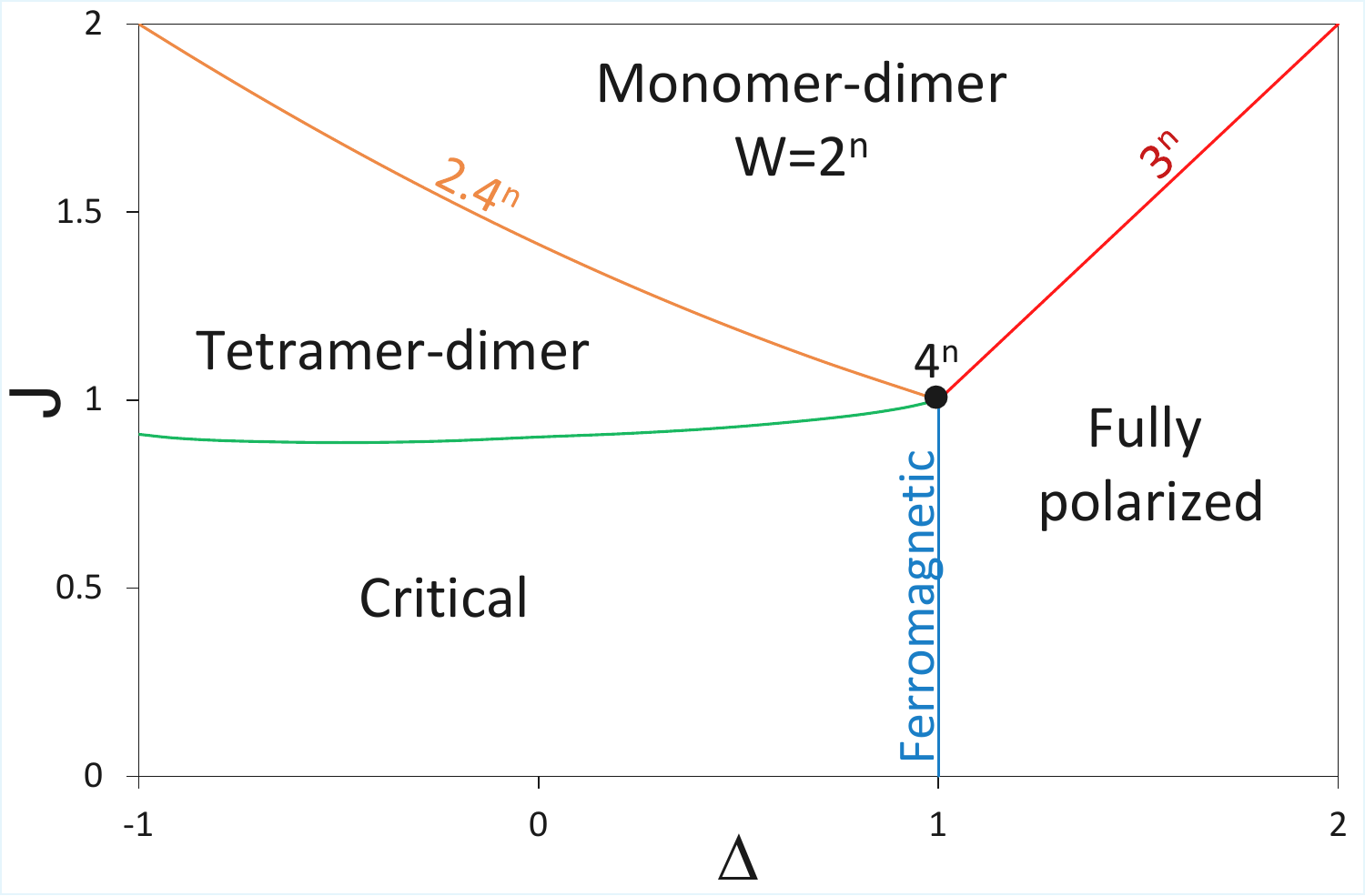}
\caption{Ground state phase diagram for the case $J_{z}=J_{\perp }=J$.
The interaction $J_0$ is taken as energy unit ($J_0=1$). }
\label{Fig_J}
\end{figure}

The phase diagram of the diamond chain with $J_{z}=J_{\perp }=J$
in the ($\Delta ,J$) plane is shown in Fig.\ref{Fig_J}. The phase
diagram consists of four ground state phases: the monomer-dimer
(MD), tetramer-dimer (TD), ferromagnetic (F), and critical (C)
phase. The latter corresponds to the ground state of the mixed
spin $(\frac{1}{2},1)$ chain with easy-plane anisotropy. All four
ground state phases meet in a quadruple point, where $J=\Delta =1$
and all phases have equal zero energies.

In the F region, the ground state is two-fold degenerate with
$S_{tot}^{z}=\pm S_{\max }^{z}$. On the line $\Delta =1$ and
$J<1$, the ground state is ferromagnetic with all possible
multiplets $-S_{\max }^{z}\leq S_{tot}^{z}\leq S_{\max }^{z}$. By
contrast, in the region (C) the ground state is non-degenerate
with $S_{tot}^{z}=0$. Lastly, the MD phase supports a $2^{n}$
degenerate ground state, composed of singlets on all diagonals and
free central spins.

The tetramer-dimer phase exists in the parameter space constrained
by $\Delta <1$ and $J_{c1}(\Delta )$ $<J<$ $J_{c2}(\Delta )$. The
boundaries $J_{c1}(\Delta )$ and $J_{c2}(\Delta )$ mark
transitions between adjacent phases, driven by energy
equivalences: between the TD and C phases at $J_{c1}(\Delta )$,
and between the TD and the MD phases at $J_{c2}(\Delta )$.

The energy of the TD state is%
\begin{equation}
E_{TD}(\Delta )=\frac{n}{2}\varepsilon _{dimer}(\Delta
)+\frac{n}{2}\varepsilon _{t}(\Delta ) \label{ETD}
\end{equation}%
where $\varepsilon _{dimer}$ is given by Eq.(\ref{Edimer}) and
$\varepsilon _{t}(\Delta )$ is the lowest energy for an isolated
diamond (tetramer) with $\mathbf{L}=1$ and $S^{z}=0$ given by
Eq.(\ref{ESz0}), which in the case $J_{z}=J_{\perp }=J$ becomes
\begin{equation}
\varepsilon _{t}(\Delta )=\frac{3}{2}\Delta
-\frac{1}{2}\sqrt{\Delta ^{2}+8}
\end{equation}

Complementarily, the ground state energy in the MD phase is
\begin{equation}
E_{MD}(\Delta )=n\varepsilon _{dimer}(\Delta )  \label{EMD}
\end{equation}

Setting $E_{TD}=E_{MD}$, we deduce the critical value $J_{c2}(\Delta )$:
\begin{equation}
J_{c2}(\Delta )=\frac{1}{2}(\sqrt{\Delta ^{2}+8}-\Delta )
\end{equation}

Analogously, $J_{c1}(\Delta )$ fulfills the condition:
\begin{equation}
E_{TD}(J_{c1},\Delta )=E_{C}(\Delta )
\end{equation}%
where $E_{C}(\Delta )$ is the ground state energy of a periodic
mixed spin-$\frac{1}{2}$ and spin-$1$ chain. Calculated
numerically for finite-sized segments $n$
(we performed DMRG calculations for $n$ up to $50$), this energy is
extrapolated to the thermodynamic limit $n\to \infty $. The
obtained dependence $J_{c1}(\Delta )$ is plotted in
Fig.\ref{Fig_J}.

The ground state phase diagram in Fig.\ref{Fig_J} is shown in the
parametric region $\Delta >-1$. In the limit $\Delta \to -1$ the
transition point $J_{c1}$ between the TD and the C phases tends to
$J\rightarrow 0.909$. For the case $\Delta =-1$ the model becomes
the isotropic AF diamond chain studied in \cite{takano1996ground}.
The ground state phase diagram of the latter model consists of the
MD, the TD and the Lieb-Mattis ferrimagnetic phase for $J<0.909$,
instead of the C phase.

Notably, all phases apart from the critical one and the quadruple
point have a finite energy gap in their excitation spectra. A
first-order phase transition occurs across all phase boundaries,
and the corresponding ground state degeneracies are analyzed in
Section IIC.

In the quadruple point the model becomes isotropic, and here the
ground state of mixed spin segments of all sizes $m_{i}$ becomes
ferromagnetic with all possible multiplets and with the energy
$\varepsilon _{0}(m_{i})=0$. The model in the isotropic point
$J=\Delta =1$ was studied in Ref.\cite{diamond1d}, where it was
shown that the ground state degeneracy is $W=4^{n}+3n-1$. This is
the maximal value of the ground state degeneracy in the F-AF
anisotropic diamond chain.

\subsection{Ground state phase diagram for diamond model with $J_{z}\neq
J_{\perp}$}

This subsection presents the ground-state phase diagram of the
model described by Hamiltonian (\ref{H}) studied within the
parametric space of anisotropy $\Delta >0$ and interactions
$J_{z}$ and $J_{\perp }$. The resulting phase diagram for
$J_{z}\neq J_{\perp }$ is qualitatively similar to the isotropic
case $J_{z}=$ $J_{\perp }$. It consists of the same four ground
state phases - the MD, the TD, the F, and the C - which all
converge at the quadruple point:
\begin{equation}
J_{\perp }=\frac{1}{\Delta },\qquad J_{z}=2\Delta -\frac{1}{\Delta }
\label{4n}
\end{equation}

\begin{figure}[tbp]
\includegraphics[width=5in,angle=0]{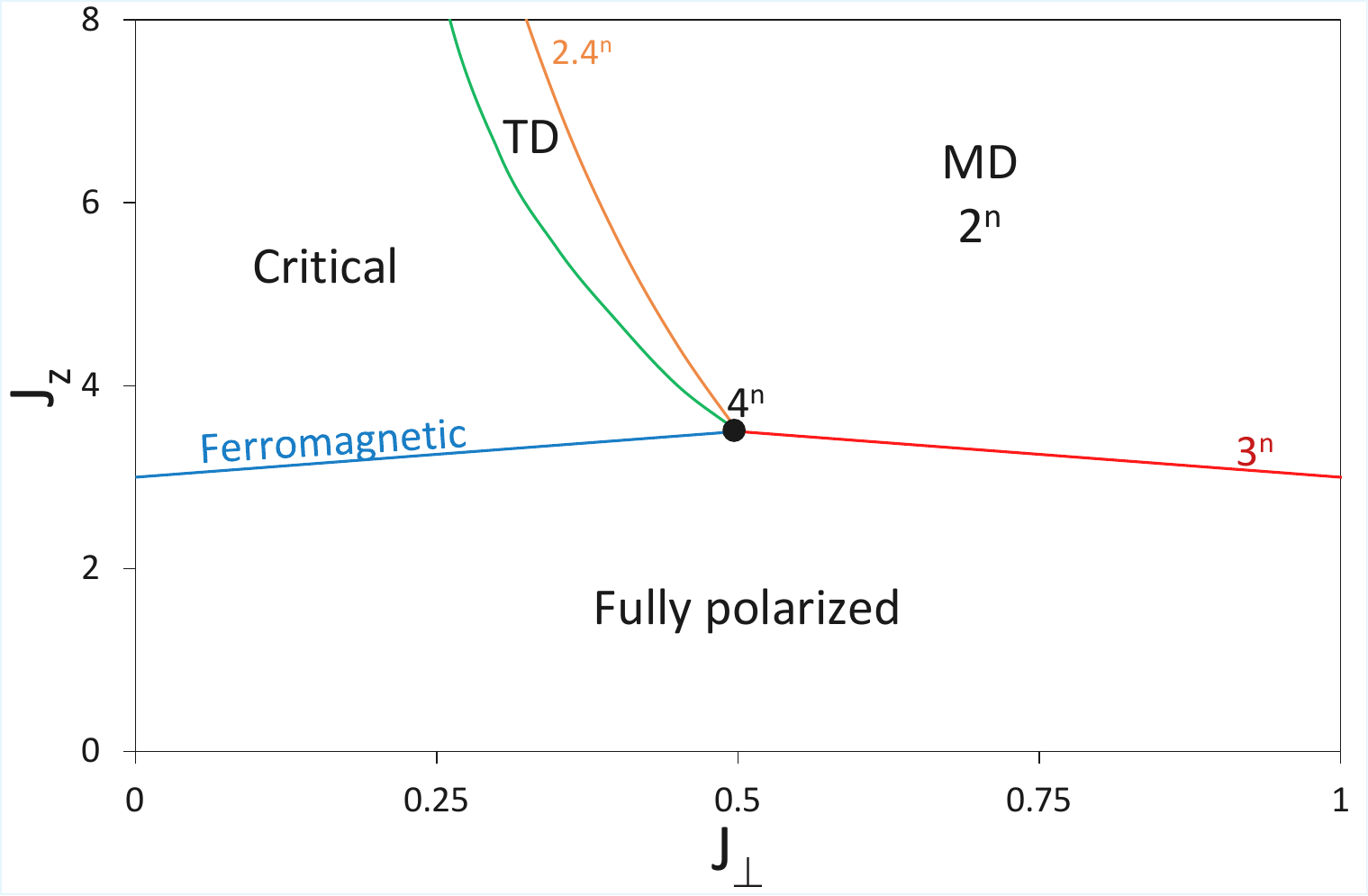}
\caption{Ground state phase diagram for the case $\Delta =2$.
The interaction $J_0$ is taken as energy unit ($J_0=1$).}
\label{Fig_Jz}
\end{figure}

As an example, the ground state phase diagram of the diamond chain
with $\Delta =2$ in plane ($J_{\perp },J_{z}$) is shown in
Fig.\ref{Fig_Jz}.

The MD phase exhibits a ground state composed of a product of
dimers and central free spins-$\frac{1}{2}$, resulting in a
degeneracy of $2^{n}$. In the TD phase, the ground state forms a
periodic structure of alternating dimers and tetramers, with the
energy $E_{TD}$ given by Eq.(\ref{ETD}). The phase boundary
between the MD and TD phases is determined by the condition
$E_{TD}=E_{MD}$, which yields the line described by
\begin{equation}
J_{z}=\frac{4}{J_{\perp }}-J_{\perp }-2\Delta  \label{TD/MD}
\end{equation}%
and valid for $J_{\perp }<\frac{1}{\Delta }$.

The boundary between the MD and the F phases is given
by the line
\begin{equation}
J_{z}=2\Delta -J_{\perp }  \label{MD/F}
\end{equation}%
for $J_{\perp }>\frac{1}{\Delta }$, where the dimer energy becomes
zero, $\varepsilon _{dimer}=0$. The ferromagnetic phase,
characterized by $S_{tot}^{z}=\pm S_{\max }^{z}$, is adjacent to
the critical phase along the line
\begin{equation}
J_{z}=J_{\perp }+2\Delta -\frac{2}{\Delta }  \label{F/C}
\end{equation}%
valid for $J_{\perp }<\frac{1}{\Delta }$. Here, the tetramer
energy is zero ($\varepsilon _{t}=0$), and the ground state
becomes ferromagnetic with all multiplets. Above this line, the
critical phase with $S_{tot}^{z}=0$ is realized. Finally, the
boundary between the critical and TD phases was determined
numerically and is plotted for $\Delta =2$ as the green line in
Fig.\ref{Fig_Jz}.

\subsection{Macroscopic degeneracy of ground state of the diamond chain}

In this subsection, we consider the anisotropic diamond chain in
the parameter space corresponding to the macroscopically
degenerate ground states found on the phase boundaries. We begin
by analyzing the degeneracy at the quadruple point, defined by
Eq.(\ref{4n}). In frustrated spin systems with competing
ferromagnetic and antiferromagnetic interactions, macroscopic
ground-state degeneracy is known to arise when the local
Hamiltonian possesses several degenerate ground states, including
the state with maximal spin \cite{KDNDR}. For the isotropic F-AF
diamond chain, this condition is satisfied when the local
Hamiltonian $\hat{H}_{i}$ has nine degenerate ground states, which
include states with $S^{z}=\pm 2$ \cite{diamond1d}. This condition
also holds for the anisotropic case. The nine ground states
comprise two with $S^{z}=\pm 2$, four with $S^{z}=\pm 1$, and
three with $S^{z}=0$. The parameters for which these nine states
form the zero-energy ($E_{i}=0$) ground state manifold of
$\hat{H}_{i}$, with all other eigenvalues $E_{i}>0$, are given by
Eq.(\ref{4n}).

Since local Hamiltonians $\hat{H}_{i}$ do not commute, the
total ground-state energy $E_{0}$ of $\hat{H}$ satisfies the
inequality:
\begin{equation}
E_{0}\geq \sum E_{i}=0  \label{ineq}
\end{equation}

The ferromagnetic state with maximal total spin $S_{\max
}^{z}=\frac{3n}{2}$ has zero energy. Consequently, the inequality
(\ref{ineq}) turns into an equality, and the ground-state energy is
exactly zero, $E_{0}=0$.

When conditions (\ref{4n}) are satisfied, the ground state of
model (\ref{H}) is macroscopically degenerate. For the ideal
diamond chain with $J_{\perp }=J_{z}$ $=J$, these conditions
coincide with those of the isotropic case $J=\Delta =1$. In this
isotropic limit, the ground state manifold consists of all
possible configurations of ferromagnetic clusters (with all
possible multiplets) separated by singlets. The number of such
ground state configurations, $W_{n}$, was derived in
\cite{diamond1d}. For a periodic chain, $W_{n}=4^{n}+3n-1$, while
for an open chain, $W_{n}=9\cdot 4^{n-1}$. This result for the open
chain implies a degeneracy equivalent to that of a system of
($n-1$) non-interacting spins-$\frac{3}{2}$ and two spins-$1$.

For the anisotropic diamond chain ($J_{\perp }\neq J_{z}$) with
parameters given by Eq.(\ref{4n}), the ground state degeneracy of
a ferromagnetic segment is identical to that of the isotropic
model. This equivalence arises because the anisotropic model
supports a set of exact eigenstates that are direct analogs of the
multiplet states in the isotropic case. To construct these states,
we consider a segment of $m$ diamonds bounded by two dimers. For
$\Delta =1$, the state with $S^{z}=S_{\max }^{z}-k$, where
$S_{\max }^{z}=\frac{3m}{2}+\frac{1}{2}$ and $k=1,\ldots 3m$, is
generated by the operator
\begin{equation}
S^{-k}=(R_{1}+R_{2})^{k}  \label{S-}
\end{equation}%
acting on the fully polarized ferromagnetic state $\left\vert
F\right\rangle$ (which has $S^z = S_{max}^z$). Here the operators
$R_1$ and $R_2$ are
\begin{eqnarray}
R_{1} &=&\sum_{i=1}^{m}(\xi _{i}^{-}+\eta _{i}^{-}) \\
R_{2} &=&\sum_{i=1}^{m+1}s_{i}^{-}  \label{R}
\end{eqnarray}

Eq.(\ref{S-}) can be rewritten in the form%
\begin{equation}
S^{-k}=\sum_{k_{i}+l_{j}=k}\prod_{k_{i}=0,1,2}^{k_{i}}(\xi _{i}^{-}+\eta
_{i}^{-})^{k_{i}}\prod_{l_{j}=0,1}s_{j}^{l_{j}}  \label{S-2}
\end{equation}

The counterpart of the lowering operator $S^-$ for the anisotropic model at the quadruple point is obtained from the exact solution:
$\tilde{S}^{-}=R_{1}+R_{2}/\Delta$. However, the higher-order operators $\tilde{S}^{-k}$ are not simple powers of $\tilde{S}^{-}$. Instead, they must be derived recursively by examining the structure of products like $\tilde{S}^{-2}$. This analysis produces the following expression for $\tilde{S}^{-k}$:
\begin{equation}
\tilde{S}^{-k}=\sum_{k_{i}+l_{j}=k}\prod_{k_{i}=0,1,2}\frac{(\xi
_{i}^{-}+\eta _{i}^{-})^{k_{i}}}{\Delta ^{k_{i}(k_{i}-1)}}\prod_{l_{j}=0,1}(%
\frac{s_{j}}{\Delta })^{l_{j}}  \label{S3}
\end{equation}

One can verify that the states $\tilde{S}^{-k}\left\vert
F\right\rangle $ are exact ground states of the ferromagnetic
segment with $S^{z}=S_{\max }^{z}-k$. Consequently, the ground
state degeneracy for the anisotropic diamond chain with parameters
$J_{\perp }=\frac{1}{\Delta }$, $J_{z}=2\Delta -\frac{1}{\Delta }$
is the same as for the isotropic case $J_{z}=$ $J_{\perp }=\Delta
=1$.

The maximal ground state degeneracy occurs when the interaction
parameters satisfy Eq.(\ref{4n}). However, a giant degeneracy
(albeit to a lesser extent) is also present on the phase boundary
between the MD and F phases, given by Eq.(\ref{MD/F}). Along this
line, there is the gap in the excitation spectrum, and the ground
state of each open segment between dimers is fully polarized, with
$S^{z}=\pm S_{max}$. The total number of degenerate ground states
is given by
\begin{equation}
W_{n}=2+\sum_{k=1}^{n}2^{k}C_{n}^{k}=3^{n}+1
\end{equation}%
where $C_{n}^{k}=\frac{n!}{k!(n-k)!}$ are the binomial
coefficients. This expression gives the ground state degeneracy on
the MD/F phase boundary in the phase diagrams of
Figs.\ref{Fig_J},\ref{Fig_Jz}.

On the phase boundary between the MD and TD phases, given by
Eq.(\ref{TD/MD}), the ground state can be described as a random
distribution of tetramers with $S^{z}=0$, each isolated from the
others by singlet dimers. For a periodic chain of $n$ diamonds,
the number of distinct configurations with $k$ such isolated
tetramers is $\frac{n}{n-k}C_{n-k}^{k}$. The number of free
spins-$\frac{1}{2}$ between neighbor singlets is $n-2k$. Summing
over all possible values of $k$ gives the total ground-state
degeneracy on the MD/TD boundary:
\begin{equation}
W=\sum_{k=0}^{n/2}\frac{n}{n-k}C_{n-k}^{k}2^{n-2k}  \label{W2.4}
\end{equation}

The maximum term in this sum is reached at
$k=\frac{\sqrt{2}-1}{2\sqrt{2}}$. Applying the saddle-point
approximation yields the asymptotic behavior:
\begin{equation}
W=\left( \frac{12}{5} \right)^{n}
\end{equation}%
which corresponds to a residual entropy per spin of $\mathcal{S}_{0}=\frac{1%
}{3n}\ln W\approx 0.292$ on the TD/MD phase boundary.

For clarity, ground-state degeneracies are annotated in the
phase diagrams in Figs. \ref{Fig_J} and \ref{Fig_Jz}. Macroscopic
degeneracy is found within the MD phase itself ($W=2^{n}$) and on
its boundaries: $W=(12/5)^{n}$ on the MD/TD boundary and
$W=3^{n}$ on the MD/F boundary. The highest degeneracy
occurs at the quadruple point, with $W=4^{n}$.

\section{F-AF anisotropic Heisenberg model on higher-dimensional diamond-decorated lattices}

The ground state phase diagram for the model on diamond-decorated
2D and 3D lattices is structurally similar to that of the diamond
chain. It comprises four phases: the monomer-dimer phase, the
tetramer-dimer phase, the ferromagnetic phase, and a critical
phase described by a mixed spin ($\frac{1}{2},1$) model on the
corresponding Lieb lattice. The boundary lines MD/TD, MD/F and F/C
are given by Eqs. (\ref{TD/MD}), (\ref{MD/F}) and (\ref{F/C}),
respectively. The quadruple point at which all four phases
converge is universally given by Eq.(\ref{4n}) for any lattice.
For the case $J_{\perp }=J_{z}=J$ the phase diagram is similar to
that of Fig.\ref{Fig_J}. At $\Delta =-1$ the model becomes the
isotropic AF Heisenberg model on diamond-decorated lattices. This
model for the square lattice was studied in
\cite{morita2016exact,hirose2016exact,hirose2018ground,caci2023phases,karl2024thermodynamic}.
The ground state phase diagram of this AF Heisenberg model
consists of the MD, the TD and the ferrimagnetic phase at
$J<0.974$ \cite{hirose2016exact}, instead of the C phase in the
F-AF model.

The ground state degeneracy at the quadruple point for the
isotropic case ($J_{\perp }=J_{z}$ $=\Delta =1$) was calculated
numerically for various 2D and 3D lattices in Ref.\cite{DKV2D}. In
this manifold, the ground states can be visualized as randomly
distributed ferromagnetic clusters of variable size and shape,
isolated from each other by diamonds with singlets on their
diagonals. The degeneracy has been computed using numerical
methods analogous to those used for bond percolation problems. For the
anisotropic model at the quadruple point, the ground state
degeneracy is identical to that of the ideal isotropic model for
all lattices. This is because the exact states given by
Eq.(\ref{S3}), which are analogous to the multiplet states of the
isotropic model, remain valid for clusters of any form and size.
Consequently, all results of \cite{DKV2D} apply directly to the
anisotropic quadruple point.

The ground state degeneracy in the MD phase is $W=2^{n}$, where
$n$ is the number of central spins $\mathbf{s}$ in the lattice. In
the TD phase, the ground state degeneracy is determined by the
number of dense dimer packings on the underlying lattice. For the
square lattice, this yields $W\sim 1.34^{n}$, corresponding to a
residual entropy per spin of $\mathcal{S}_{0}=0.05831$ and for the
hexagonal lattice $\mathcal{S}_{0}=0.0422$ \cite{morita2016exact}.
The critical phase on 2D and 3D lattices, modeled by the
mixed-spin system on a Lieb lattice, has a non-degenerate ground
state with $S^{z}=0$. The fully polarized ferromagnetic phase is
two-fold degenerate for any lattice.

The phase boundaries involving the C phase (TD/C and C/F) do not
exhibit macroscopic ground state degeneracy. In contrast, the
degeneracy on the TD/MD boundary maps to the problem of counting
all possible configurations of non-touching dimers on the lattice,
resulting in a higher degeneracy than in the adjacent TD and MD
phases. Our calculations of the degeneracy on the TD/MD boundary
based on the transfer-matrix approach \cite{lieb2004solution} give
$W\approx 2.704^{n}$ and the corresponding residual entropy
$\mathcal{S}_{0}=0.199$.

On the MD/F boundary, the ground state manifold consists of
randomly distributed ferromagnetic clusters, similar to the
isotropic case studied in \cite{DKV2D}, but with a key
modification that substantially reduces the degeneracy. The
difference lies in the contribution of each cluster: in the
isotropic case all multiplets of the ferromagnetic state
contribute, whereas on the MD/F boundary, only the two fully
polarized states with $S^{z}=\pm S_{\max }^{z}$ are ground states
for any cluster. In the following, we briefly outline the method
for calculating this ground state degeneracy and present the
results.

\subsection{Ground state degeneracy on the MD/F phase boundary}

Each singlet on a diamond diagonal (see Fig.\ref{Fig_diamond})
effectively breaks the bond between the corresponding spins
$\mathbf{s}_{\mathbf{i}}$ and $\mathbf{s}_{\mathbf{j}}$.
Consequently, the ground state degeneracy can be computed by
enumerating all possible configurations of diamonds with singlet
or triplet diagonals. Each such configuration corresponds to a set
of exact ground states of the total Hamiltonian (\ref{H}).
Therefore, the total degeneracy is the sum of the degeneracies
associated with every possible configuration of singlet diagonals.
This formulation maps the problem directly onto a \textit{bond
percolation problem}. Here, a diamond with a triplet diagonal
represents a connected bond, while a diamond with a singlet
diagonal represents a disconnected bond. A similar mapping
of the ground-state degeneracy in a percolation problem was
previously applied to electron systems \cite{Pauli-Correlated}.
On the other hand, the partition function considered here is formally
identical to that of an Ising model expressed in terms of
Fortuin-Kasteleyn clusters \cite{MHB}.

For a given configuration $\omega _{K}$ containing $K$ triplet
diagonals (connected bonds), the lattice partitions into multiple
disconnected clusters. The ground state of each cluster consists
of the two fully polarized states with $S^{z}=\pm S_{\max }^{z}$,
leading to a degeneracy of $2$, independent of the cluster's size
or shape. The total ground state degeneracy for the configuration
$\omega _{K}$ is the product of the degeneracies of all its
incoming clusters
\begin{equation}
W(\omega _{K},n)=2^{m}  \label{W}
\end{equation}%
where $m$ is the number of clusters in the configuration $\omega _{K}$.

\begin{figure}[tbp]
\includegraphics[width=5in,angle=0]{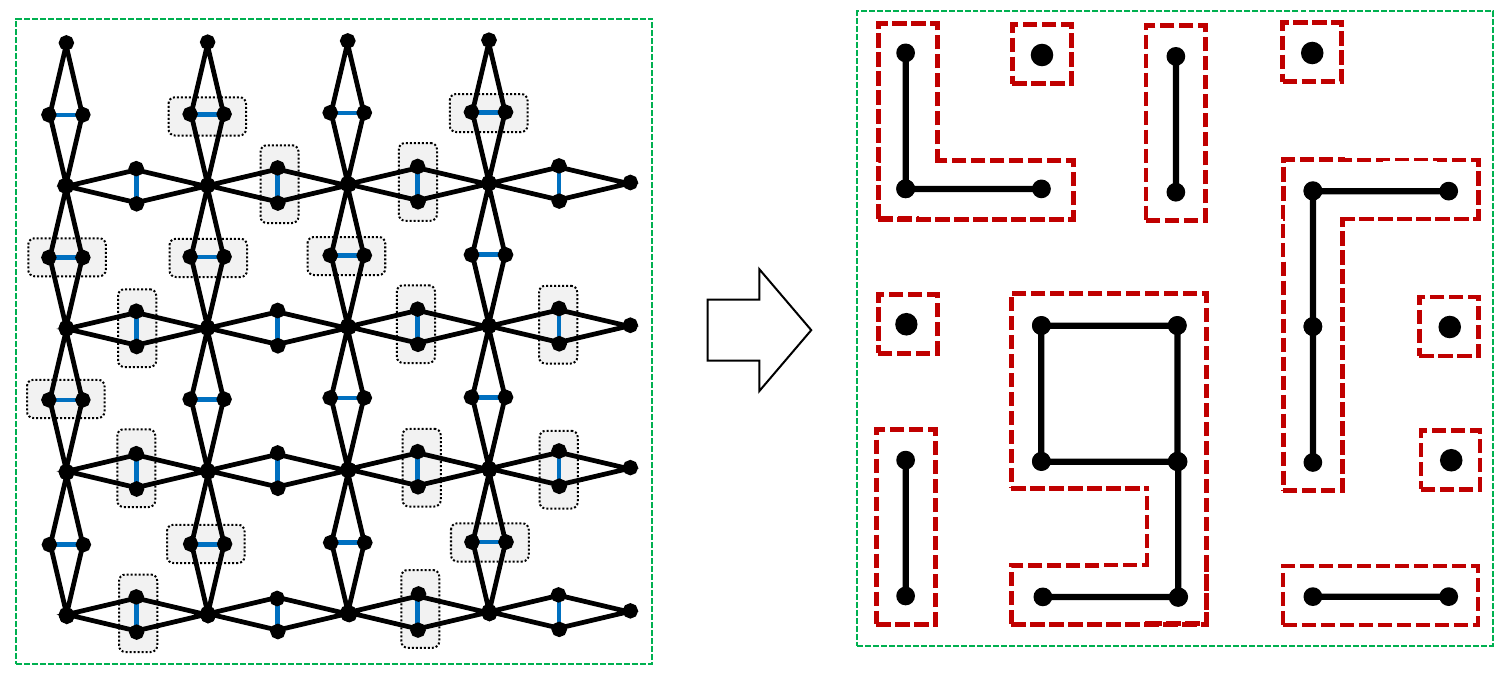}
\caption{Ideal diamond decorated square lattice 4x4 with particular
configuration of diamonds with singlets on diagonal (shaded diagonals) and
the corresponding percolation configuration with connected and disconnected
bonds.}
\label{Fig_percolation}
\end{figure}

To illustrate the mapping to the percolation problem, we examine a
$4\times 4$ sample of the square lattice with open boundary
conditions, depicted in Fig. \ref{Fig_percolation}. The left panel
shows a specific configuration of singlet diagonals (represented
by shaded rectangles) in the original spin model. The
corresponding percolation framework is shown in the right panel,
where diamonds with singlet diagonals are removed (disconnected
bonds) and those with triplet diagonals are replaced by connected
bonds. In this configuration, eleven distinct clusters of varying
sizes emerge, resulting in a degeneracy of $W=2^{11}$.

To calculate the total number of ground states $W(K,n)$ for a
fixed number of connected bonds $K$ on a lattice of $n$ sites
(central spins), we sum the degeneracy $W(\omega _{K},n)$ over all
configurations $\omega _{K}$ with exactly $K$ connected bonds:
\begin{equation}
W(K,n)=\sum_{\omega _{K}}W(\omega _{K},n)  \label{ZK}
\end{equation}

The total ground-state degeneracy $W(n)$ is then obtained by
summing over all possible values of $K$:
\begin{equation}
W(n)=\sum_{K=0}^{N_{b}}W(K,n)  \label{Z}
\end{equation}%
where $N_{b}$ is the total number of bonds.

\begin{figure}[tbp]
\includegraphics[width=5in,angle=0]{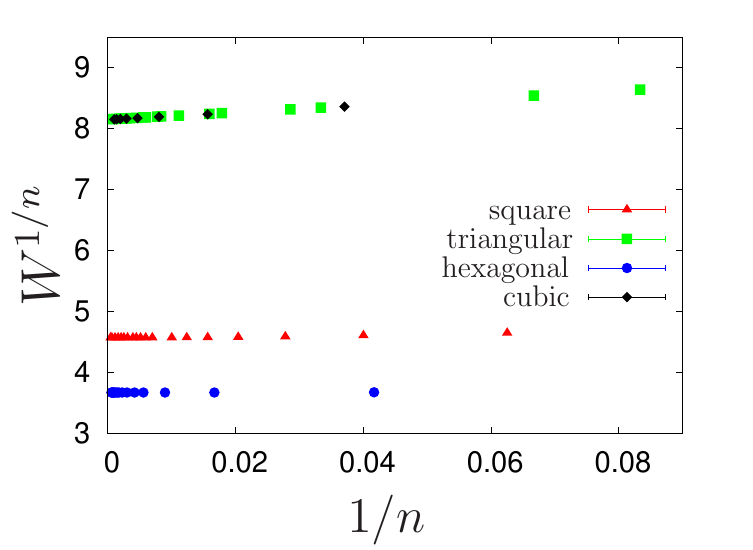}
\caption{The values of $G=W^{1/n}$ as a function of $1/n$ for the
square, triangular, honeycomb and cubic lattices.} \label{fig:zn}
\end{figure}

Details of the numerical computation of $W(n$) are provided in
Ref.\cite{DKV2D}. The results show that for all the studied lattices
(hexagonal, square, triangular, cubic), the ground-state
degeneracy grows exponentially with $n$, $W=G^{n}$, with the value
$G$ depending on the lattice. The finite-size scaling of
$G=W^{1/n}$ versus $1/n$ is shown in Fig. \ref{fig:zn}.

As illustrated in Fig. \ref{fig:zn}, the values of $G$ converge to
a finite limit as $n\to\infty $ for each lattice, giving the
thermodynamic value of $G$. The residual entropy per spin
$\mathcal{S}_{0}=\frac{\ln G}{z+1}$($z$ is coordination number of
the lattice), for all studied lattices are presented in Table 1,
alongside the corresponding values for the diamond models at the
quadruple point. As shown in Table 1, the residual entropy on the
MD/F boundary is lower than at the quadruple point for all 2D and
3D lattices studied.

\begin{table}[tbp]
\caption{The values of residual entropy per spin,
$\mathcal{S}_{0}$, on the MD/F boundary and at the quadruple point
for different lattices}
\begin{ruledtabular}
\begin{tabular}{ccc}
Lattice & MD/F boundary & Quadruple point    \\
\hline Chain & $0.366$ & $0.462$ \\
Hexagonal & $0.326$ & $0.402$ \\
Square & $0.304$ & $0.362$ \\
Triangular  & $0.3$ & $0.314$ \\
Cubic &  $0.3$ & $0.302$ \\
\end{tabular}
\end{ruledtabular}
\end{table}

\begin{figure}[tbp]
\includegraphics[width=5in,angle=0]{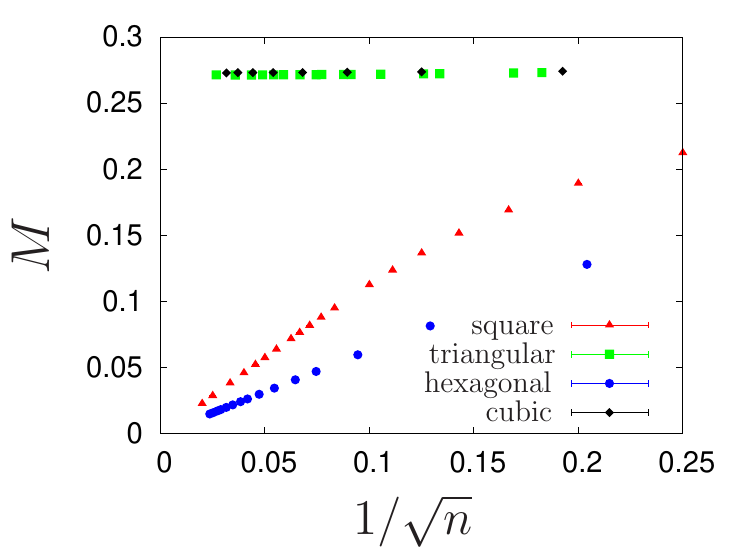}
\caption{Magnetization per spin as a function of $n^{-1/2}$ for
square, honeycomb, triangular and cubic lattices.} \label{fig:M}
\end{figure}

The magnetization for anisotropic models on the MD/F boundary can
be calculated using the method from Ref.\cite{DKV2D}. The results
in Fig. \ref{fig:M} show that the ground-state magnetization
vanishes in the thermodynamic limit ($n\to\infty$) for hexagonal
and square lattices, but remains finite for triangular and cubic
lattices. This behavior is consistent with the results for the
quadruple point reported in \cite{DKV2D}.

\section{Summary}

We have studied the ground state properties of the anisotropic
spin-$\frac{1}{2}$ Heisenberg model composed of ideal diamond
units with competing ferro- and antiferromagnetic interactions.
The study encompasses both one-dimensional diamond chains and
their higher-dimensional generalizations on diamond-decorated
square, hexagonal, triangular, and cubic lattices.

For the one-dimensional diamond chain, we established the complete
ground state phase diagram in the parameter space ($\Delta $, $J_{\perp }$,
$J_{z}$). This diagram consists in four phases:
the ferromagnetic phase; the gapless, non-magnetic critical phase;
the monomer-dimer phase with a $2^n$-fold degenerate ground state
of decoupled spins and singlets; and the tetramer-dimer phase,
characterized by a periodic arrangement of singlets and tetramers.
All phases excluding the C phase exhibit a finite energy gap in
the excitation spectrum, and first-order quantum phase transitions
occur across all phase boundaries. These four phases converge at a
quadruple point, where the model has the maximal ground state
degeneracy $W\simeq 4^n$ and the finite residual entropy.
Significant macroscopic degeneracy, though to a lesser extent, is
also found within the MD phase ($W=2^n$) and on its boundaries
with the F phase ($W\simeq 3^n$) and TD phase ($W\simeq(12/5)^n$).

The ground state phase diagram of two- and three-dimension
lattices has a qualitatively similar structure - featuring the
same four phases and the quadruple point. Macroscopic ground
state degeneracy exists in the MD and TD phases, and on the
MD/TD and MD/F phase boundaries. The maximal degeneracy is
achieved in the quadruple point and the corresponding values
$W(n)$ for different lattices were obtained in Ref.\cite{DKV2D}.
The ground state degeneracy on the MD/TD boundary for the square
lattice was calculated using transfer-matrix method. To compute
the ground state degeneracy on the MD/F boundary, we use a mapping
to a classical bond percolation problem. In this framework, a
diamond with a triplet (singlet) diagonal corresponds to a
connected (disconnected) bond. The total ground state degeneracy
is then obtained by summing over all percolation configurations,
with each resulting isolated cluster contributing a degeneracy
factor specific to the phase. Using this mapping in conjunction
with numerical Monte Carlo techniques, we confirm that the
degeneracy grows exponentially with the system size $W(n)\sim G^{n}$,
for all the studied lattices.

The values of the residual entropy $\mathcal{S}_{0}$, which depend
on the lattice geometry, are presented in Table 1. These results
show a consistent trend: the residual entropy on the MD/F boundary
is lower than at the quadruple point. In addition, we calculated
the magnetization on the MD/F boundary, revealing a distinct
lattice dependence: it vanishes in the thermodynamic limit for
hexagonal and square lattices, but remains finite for triangular
and cubic lattices. This behavior mirrors that observed at the
isotropic quadruple point.

\begin{acknowledgments}
The present work was funded by the Ministry of Science and Higher
Education, Russian Federation (Research theme state registration
number 125020401357-4).
\end{acknowledgments}

\bibliography{diamond_anis}

@article{DKV2D,
  title = {Macroscopic ground state degeneracy of the Heisenberg model with ferromagnetic and antiferromagnetic interactions on diamond-decorated lattices},
  author = {Dmitriev, D. V. and Krivnov, V. Ya. and Vasilyev, O. A.},
  journal = {Phys. Rev. B},
  volume = {112},
  issue = {9},
  pages = {094426},
  numpages = {15},
  year = {2025},
  month = {Sep},
  publisher = {American Physical Society},
  doi = {10.1103/drjr-wt6n},
  url = {https://link.aps.org/doi/10.1103/drjr-wt6n}
}

@book{diep,
  title={Frustrated spin systems},
  author={Diep, Hung T and others},
  year={2013},
  publisher={World scientific}
}

@book{mila,
  title={Introduction to frustrated magnetism: materials, experiments, theory},
  author={Lacroix, Claudine and Mendels, Philippe and Mila, Fr{\'e}d{\'e}ric},
  volume={164},
  year={2011},
  publisher={Springer Science \& Business Media}
}

@article{flat,
  title={Strongly correlated flat-band systems: The route from Heisenberg spins to Hubbard electrons},
  author={Derzhko, Oleg and Richter, Johannes and Maksymenko, Mykola},
  journal={International Journal of Modern Physics B},
  volume={29},
  number={12},
  pages={1530007},
  year={2015},
  publisher={World Scientific}
}

@article{shulen,
  title = {Magnetic-Field Induced Spin-Peierls Instability in Strongly Frustrated Quantum Spin Lattices},
  author = {Richter, Johannes and Derzhko, Oleg and Schulenburg, J\"org},
  journal = {Physical review letters},
  volume = {93},
  issue = {10},
  pages = {107206},
  numpages = {4},
  year = {2004},
  month = {Sep},
  publisher = {American Physical Society},
}

@article{schmidt,
  title={Independent magnon states on magnetic polytopes},
  author={Schnack, J{\"u}rgen and Schmidt, H-J and Richter, Johannes and Schulenburg, J{\"o}rg},
  journal={The European Physical Journal B},
  volume={24},
  pages={475--481},
  year={2001},
  publisher={Springer}
}

@article{honecker,
  title={Exact eigenstates and macroscopic magnetization jumps in strongly frustrated spin lattices},
  author={Richter, Johannes and Schulenburg, J{\"o}rg and Honecker, Andreas and Schnack, J{\"u}rgen and Schmidt, Heinz-J{\"u}rgen},
  journal={Journal of Physics: Condensed Matter},
  volume={16},
  number={11},
  pages={S779},
  year={2004},
  publisher={IOP Publishing}
}

@article{Derzhko,
  title={Finite low-temperature entropy of some strongly frustrated quantum spin lattices in the vicinity of the saturation field},
  author={Derzhko, Oleg and Richter, Johannes},
  journal={Physical Review B},
  volume={70},
  number={10},
  pages={104415},
  year={2004},
  publisher={APS}
}

@article{zhitomir,
  title={Magnetocaloric effect in one-dimensional antiferromagnets},
  author={Zhitomirsky, ME and Honecker, Andreas},
  journal={Journal of Statistical Mechanics: Theory and Experiment},
  volume={2004},
  number={07},
  pages={P07012},
  year={2004},
  publisher={IOP Publishing}
}

@article{KDNDR,
  title = {Delta chain with ferromagnetic and antiferromagnetic interactions at the critical point},
  author = {Krivnov, V. Ya. and Dmitriev, D. V. and Nishimoto, S. and Drechsler, S.-L. and Richter, J.},
  journal = {Phys. Rev. B},
  volume = {90},
  issue = {1},
  pages = {014441},
  numpages = {11},
  year = {2014},
  month = {Jul},
  publisher = {American Physical Society},
}

@article{DKRS,
  title={Thermodynamics of a delta chain with ferromagnetic and antiferromagnetic interactions},
  author={Dmitriev, D V and Krivnov, V Ya and Richter, Johannes and Schnack, J{\"u}rgen},
  journal={Physical Review B},
  volume={99},
  number={9},
  pages={094410},
  year={2019},
  publisher={APS}
}

@article{DKRS2,
  title={Exact magnetic properties for classical delta-chains with ferromagnetic and antiferromagnetic interactions in applied magnetic field},
  author={Dmitriev, D V and Krivnov, V Ya and Schnack, J{\"u}rgen and Richter, Johannes},
  journal={Physical Review B},
  volume={101},
  number={5},
  pages={054427},
  year={2020},
  publisher={APS}
}

@article{DKRS3,
  title={Flat-band physics in the spin-1/2 sawtooth chain},
  author={Derzhko, Oleg and Schnack, J{\"u}rgen and Dmitriev, Dmitry V and Krivnov, Valery Ya and Richter, Johannes},
  journal={The European Physical Journal B},
  volume={93},
  pages={1--12},
  year={2020},
  publisher={Springer}
}

@article{alum,
  title={Spin-liquid ground state in the spin 1/2 distorted diamond chain compound ${K}_3{Cu}_3{Al}{O}_2 {({SO}_4)}_4$},
  author={Fujihala, Masayoshi and Koorikawa, Hiroko and Mitsuda, Setsuo and Hagihala, Masato and Morodomi, Hiroki and Kawae, Tatsuya and Matsuo, Akira and Kindo, Koichi},
  journal={Journal of the Physical Society of Japan},
  volume={84},
  number={7},
  pages={073702},
  year={2015},
  publisher={The Physical Society of Japan}
}

@article{anis3,
  title={Two-dimensional spin models with macroscopic degeneracy},
  author={Dmitriev, D V and Krivnov, V Ya},
  journal={Journal of Physics: Condensed Matter},
  volume={33},
  number={43},
  pages={435802},
  year={2021},
  publisher={IOP Publishing}
}

@article{derzhko2007universal,
  title={Universal properties of highly frustrated quantum magnets in strong magnetic fields},
  author={Derzhko, Oleg and Richter, Johannes and Honecker, Andreas and Schmidt, H-J},
  journal={Low Temperature Physics},
  volume={33},
  number={9},
  pages={745--756},
  year={2007},
  publisher={AIP Publishing}
}

@article{schulenburg2002macroscopic,
  title={Macroscopic magnetization jumps due to independent magnons in frustrated quantum spin lattices},
  author={Schulenburg, J{\"o}rg and Honecker, Andreas and Schnack, J{\"u}rgen and Richter, Johannes and Schmidt, H-J},
  journal={Physical review letters},
  volume={88},
  number={16},
  pages={167207},
  year={2002},
  publisher={APS}
}

@article{morita2016exact,
  title={Exact nonmagnetic ground state and residual entropy of s=1/2 Heisenberg diamond spin lattices},
  author={Morita, Katsuhiro and Shibata, Naokazu},
  journal={Journal of the Physical Society of Japan},
  volume={85},
  number={3},
  pages={033705},
  year={2016},
  publisher={The Physical Society of Japan}
}

@article{hirose2016exact,
  title={Exact realization of a quantum-dimer model in Heisenberg antiferromagnets on a diamond-like decorated lattice},
  author={Hirose, Yuhei and Oguchi, Akihide and Fukumoto, Yoshiyuki},
  journal={Journal of the Physical Society of Japan},
  volume={85},
  number={9},
  pages={094002},
  year={2016},
  publisher={The Physical Society of Japan}
}

@article{hirose2018ground,
  title={Ground-state properties of spin-1/2 Heisenberg antiferromagnets with frustration on the diamond-like-decorated square and triangular lattices},
  author={Hirose, Yuhei and Miura, Shoma and Yasuda, Chitoshi and Fukumoto, Yoshiyuki},
  journal={AIP Advances},
  volume={8},
  number={10},
  pages={101427},
  year={2018},
  publisher={AIP Publishing}
}

@article{caci2023phases,
  title={Phases of the spin-1/2 Heisenberg antiferromagnet on the diamond-decorated square lattice in a magnetic field},
  author={Caci, Nils and Karl'ov{\'a}, Katar{\'\i}na and Verkholyak, Taras and Stre{\v{c}}ka, Jozef and Wessel, Stefan and Honecker, Andreas},
  journal={Physical Review B},
  volume={107},
  number={11},
  pages={115143},
  year={2023},
  publisher={APS}
}

@article{karl2024thermodynamic,
  title={Thermodynamic properties of the macroscopically degenerate tetramer-dimer phase of the spin-1/2 Heisenberg model on the diamond-decorated square lattice},
  author={Karl'ov{\'a}, Katar{\'\i}na and Honecker, Andreas and Caci, Nils and Wessel, Stefan and Stre{\v{c}}ka, Jozef and Verkholyak, Taras},
  journal={Physical Review B},
  volume={110},
  number={21},
  pages={214429},
  year={2024},
  publisher={APS}
}

@article{diamond1d,
  title = {Macroscopic degeneracy of the ground state in the frustrated Heisenberg diamond chain},
  author = {Dmitriev, D. V. and Krivnov, V. Ya.},
  journal = {Physical Review B},
  volume = {111},
  issue = {6},
  pages = {064427},
  numpages = {11},
  year = {2025},
  month = {Feb},
  publisher = {American Physical Society},
}

@article{diep2,
     author = {Hung The Diep},
     title = {Frustrated spin systems: history of the emergence of a modern physics},
     journal = {Comptes Rendus. Physique},
     pages = {225--251},
     publisher = {Acad\'emie des sciences, Paris},
     volume = {26},
     year = {2025},
}

@article{hon2,
  title={Magnetocaloric effect in two-dimensional spin-1/2 antiferromagnets},
  author={Honecker, A and Wessel, S},
  journal={Physica B: Condensed Matter},
  volume={378},
  pages={1098--1099},
  year={2006},
  publisher={Elsevier}
}

@article{zhitomirsky2005high,
  title={High field properties of geometrically frustrated magnets},
  author={Zhitomirsky, ME and Tsunetsugu, Hirokazu},
  journal={Progress of Theoretical Physics Supplement},
  volume={160},
  pages={361--382},
  year={2005},
  publisher={Oxford Academic}
}

@article{derzhko2006universal,
  title={Universal low-temperature behavior of frustrated quantum antiferromagnets in the vicinity of the saturation field},
  author={Derzhko, Oleg and Richter, Johannes},
  journal={The European Physical Journal B-Condensed Matter and Complex Systems},
  volume={52},
  pages={23--36},
  year={2006},
  publisher={Springer}
}

@article{richter2018thermodynamic,
  title={Thermodynamic properties of ${Ba}_{2} {Co} {Si}_{2} {O}_{6} {Cl}_{2}$ in a strong magnetic field: Realization of flat-band physics in a highly frustrated quantum magnet},
  author={Richter, Johannes and Krupnitska, Olesia and Baliha, Vasyl and Krokhmalskii, Taras and Derzhko, Oleg},
  journal={Physical Review B},
  volume={97},
  number={2},
  pages={024405},
  year={2018},
  publisher={APS}
}

@article{zhitomirsky2003enhanced,
  title={Enhanced magnetocaloric effect in frustrated magnets},
  author={Zhitomirsky, ME},
  journal={Physical Review B},
  volume={67},
  number={10},
  pages={104421},
  year={2003},
  publisher={APS}
}

@article{PhysRevB.111.144420,
  title = {Spin dynamics and 1/3 magnetization plateau in the coupled distorted diamond chain compound ${K}_{2}{Cu}_{3}{({MoO}_{4})}_{4}$},
  author = {Murugan, G. Senthil and Khatua, J. and Kim, Suyoung and Mun, Eundeok and Babu, K. Ramesh and Kim, Heung-Sik and Huang, C.-L. and Kalaivanan, R. and Kumar, U. Rajesh and Muthuselvam, I. Panneer and Chen, W. T. and Krishnamoorthi, Sritharan and Choi, K.-Y. and Sankar, R.},
  journal = {Phys. Rev. B},
  volume = {111},
  issue = {14},
  pages = {144420},
  numpages = {13},
  year = {2025},
  month = {Apr},
  publisher = {American Physical Society},
  doi = {10.1103/PhysRevB.111.144420},
  url = {https://link.aps.org/doi/10.1103/PhysRevB.111.144420}
}

@article{HONG20043271,
title = {Cyano-bridged ${Fe}$(II)–${Cu}$(II) bimetallic assemblies: honeycomb-like and pentanuclear structures},
journal = {Inorganica Chimica Acta},
volume = {357},
number = {11},
pages = {3271-3278},
year = {2004},
issn = {0020-1693},
doi = {https://doi.org/10.1016/j.ica.2004.04.004},
url = {https://www.sciencedirect.com/science/article/pii/S0020169304001963},
author = {Chang Seop Hong and Young Sin You},
}

@article{takano1996ground,
  title={Ground states with cluster structures in a frustrated Heisenberg chain},
  author={Takano, K and Kubo, K and Sakamoto, H},
  journal={Journal of Physics: Condensed Matter},
  volume={8},
  number={35},
  pages={6405},
  year={1996},
  publisher={IOP Publishing}
}

@article{alcaraz1997critical,
  title={Critical behaviour of mixed Heisenberg chains},
  author={Alcaraz, FC and Malvezzi, AL},
  journal={Journal of Physics A: Mathematical and General},
  volume={30},
  number={3},
  pages={767},
  year={1997},
  publisher={IOP Publishing}
}

@incollection{lieb2004solution,
  title={Solution of the dimer problem by the transfer matrix method},
  author={Lieb, Elliott H},
  booktitle={Condensed Matter Physics and Exactly Soluble Models: Selecta of Elliott H. Lieb},
  pages={537--539},
  year={2004},
  publisher={Springer}
}

@article{Pauli-Correlated,
  title = {Flat-Band Ferromagnetism as a Pauli-Correlated Percolation Problem},
  author = {Maksymenko, M. and Honecker, A. and Moessner, R. and Richter, J. and Derzhko, O.},
  journal = {Phys. Rev. Lett.},
  volume = {109},
  issue = {9},
  pages = {096404},
  numpages = {5},
  year = {2012},
  month = {Aug},
  publisher = {American Physical Society},
  doi = {10.1103/PhysRevLett.109.096404},
  url = {https://link.aps.org/doi/10.1103/PhysRevLett.109.096404}
}

@article{MHB,
  title={Monte Carlo study of the Ising model phase transition in terms of the percolation transition of “physical clusters”},
  author={De Meo, Marco D'Onorio and Heermann, Dieter W and Binder, Kurt},
  journal={Journal of statistical physics},
  volume={60},
  number={5},
  pages={585--618},
  year={1990},
  publisher={Springer}
}

\end{document}